%% file: main.tex
\documentclass[ALICE,manyauthors]{cernphprep}

\usepackage[comma,square,numbers,sort&compress]{natbib}
\usepackage{hyperref}
\usepackage{lineno}
\usepackage{xcolor}
\usepackage{textcomp}
\usepackage[T1]{fontenc}

\input{commands}

\begin{document}%

\begin{titlepage}
\PHyear{2021}
\PHnumber{139}      
\PHdate{14 July}  
%

\title{Hypertriton production in p--Pb collisions at \snn= 5.02 TeV}
\ShortTitle{Hypertriton production in p--Pb collisions at the \snn=5.02 TeV}   
\hyphenation{nu-cleo-syn-the-sis}
\Collaboration{ALICE Collaboration\thanks{See Appendix~\ref{app:collab} for the list of collaboration members}}
\ShortAuthor{ALICE Collaboration} 

\begin{abstract}
\input{abstract}
\end{abstract}
\end{titlepage}
\setcounter{page}{2}

\input{text}

\newenvironment{acknowledgement}{\relax}{\relax}
\begin{acknowledgement}
\section*{Acknowledgements}
\input{fa_2021-07-06.tex}    
\end{acknowledgement}

\bibliographystyle{utphys}   
\bibliography{biblio}

\newpage

\appendix
\section{The ALICE Collaboration}
\label{app:collab}
\input{Alice_Authorlist_2021-07-06.tex}  
\end{document}

%% file: commands.tex
\usepackage{xspace}

\newcommand{\PbPb}         {\mbox{Pb--Pb}\xspace}

\newcommand{\pPb}          {\mbox{p--Pb}\xspace}

\newcommand{\snn}          {\ensuremath{\sqrt{s_{\mathrm{NN}}}}\xspace}
\newcommand{\pt}           {\ensuremath{p_{\rm T}}\xspace}

\newcommand{\dEdx}         {\ensuremath{\textrm{d}E/\textrm{d}x}\xspace}

\newcommand{\dVdy}         {\ensuremath{\mathrm{d}V/\mathrm{d}y}\xspace}

\newcommand{\nineH}        {$\sqrt{s}~=~0.9$~Te\kern-.1emV\xspace}
\newcommand{\seven}        {$\sqrt{s}~=~7$~Te\kern-.1emV\xspace}
\newcommand{\twoH}         {$\sqrt{s}~=~0.2$~Te\kern-.1emV\xspace}
\newcommand{\twosevensix}  {$\sqrt{s}~=~2.76$~Te\kern-.1emV\xspace}
\newcommand{\five}         {$\sqrt{s}~=~5.02$~Te\kern-.1emV\xspace}
\newcommand{\twosevensixnn}{$\sqrt{s_{\mathrm{NN}}}~=~2.76$~Te\kern-.1emV\xspace}
\newcommand{\fivenn}       {$\sqrt{s_{\mathrm{NN}}}~=~5.02$~Te\kern-.1emV\xspace}

\newcommand{\GeVc}         {Ge\kern-.1emV/$c$\xspace}
\newcommand{\MeVc}         {Me\kern-.1emV/$c$\xspace}
\newcommand{\TeV}          {Te\kern-.1emV\xspace}
\newcommand{\GeV}          {Ge\kern-.1emV\xspace}
\newcommand{\MeV}          {Me\kern-.1emV\xspace}
\newcommand{\GeVmass}      {Ge\kern-.2emV/$c^2$\xspace}
\newcommand{\MeVmass}      {Me\kern-.2emV/$c^2$\xspace}

\newcommand{\Hee}          {\ensuremath{^{3}\mathrm{He}}\xspace}
\newcommand{\antiHee}      {\ensuremath{^{3}\overline{\mathrm{He}}}\xspace}
\newcommand{\hyp}          {\ensuremath{^{3}_{\Lambda}\mathrm H}\xspace}
\newcommand{\antihyp}      {\ensuremath{^{3}_{\overline{\Lambda}} \overline{\mathrm H}}\xspace}
\newcommand{\he}           {\ensuremath{^3\mathrm{He}}\xspace}

\newcommand{\dd}     {\ensuremath{\mathrm{d}}\xspace}

%% file: abstract.tex
The study of nuclei and antinuclei production has proven to be a powerful tool to investigate the formation mechanism of loosely bound states in high-energy hadronic collisions. The first measurement of the production of $\text{\hyp}$ in \pPb collisions at \snn = 5.02 TeV is presented in this Letter. Its production yield measured in the rapidity interval $-1 < y < 0$ for the 40\% highest multiplicity \pPb collisions is $\dd N /\dd y =[\mathrm{6.3 \pm 1.8 (stat.) \pm 1.2 (syst.) ] \times 10^{-7}}$.
The measurement is compared with the expectations of statistical hadronisation and coalescence models, which describe the nucleosynthesis in hadronic collisions. These two models predict very different yields
of the hypertriton in charged particle multiplicity environments relevant to small
collision systems such as \pPb and therefore the measurement of $\dd N /\dd y$ is crucial to distinguish between them.
The precision of this measurement leads to the exclusion with a significance larger than 6.9$\sigma$ of some configurations of the statistical hadronization model, thus constraining the theory behind the production of loosely bound states at hadron colliders.

%% file: text.tex
In the last few decades, the production of deuterons, $^3$H, \Hee, $\mathrm{^{4}He}$ and their charge conjugates was measured in many different colliding systems and energies.
The results of the measurements in hadronic and heavy-ion collisions at the LHC~\cite{Adam:2015vda,Acharya:2017dmc,Acharya:2019rgc,Acharya:2019rys,Acharya:2019xmu,Acharya:2020sfy,Acharya:2017bso},
in e$^+$e$^-$ collisions at LEP~\cite{Schael:2006fd}, at
lower-energy collider~\cite{Alper:1973my,Henning:1977mt,Alexopoulos:2000jk,Adler:2001uy,Adler:2004uy,Aktas:2004pq,Asner:2006pw,Agakishiev:2011ib} and fixed target experiments~\cite{SimonGillo:1995dh,Armstrong:2000gd,Afanasev:2000ku,Anticic:2004yj} significantly constrained the parameter space for production models like coalescence~\cite{Mrowczynski:1987oid,Scheibl:1998tk,Sun:2018mqq} and statistical hadronisation~\cite{Andronic:2010qu,Vovchenko:2018fiy}, yet they were unable to decisively  discriminate between these two models.
The interest in the phenomenon of nucleosynthesis in the final state of hadronic collisions has risen again in recent years owing to its relevance in dark matter searches in space~\cite{Blum:2017qnn, Korsmeier:2017xzj}.
A precise modelling of the production of nuclei and antinuclei is required for the interpretation of the expected fluxes of antinuclei originating from dark matter annihilation, and for the relevant Standard Model background channels.

For large colliding systems, such as Pb--Pb collisions at the LHC, the predictions of statistical hadronisation and coalescence models are very similar and they are both able to describe the measured production of nuclei~\cite{Bellini:2018epz}. The statistical hadronisation model (SHM) describes the system as a hadron-resonance gas (HRG) in thermal equilibrium at hadron emission, hence it predicts particle yields starting from the volume and the temperature of the system at chemical freeze-out ($T_{\mathrm{chem}}$).
The Grand Canonical formulation of the SHM describes the measured production yields of light hadrons and nuclei in Pb--Pb collisions at 2.76 TeV with $T_{\mathrm{chem}}$ = 155 MeV~\cite{Acharya:2019xmu}. This temperature, which successfully describes the yield of light hadrons in central \PbPb collisions, is one to two orders of magnitude larger than the typical binding energies of light nuclei (a few MeVs), and nuclei are likely to interact with the other hadrons in the dense HRG after chemical  freeze-out due to the large cross sections~\cite{Oliinychenko:2018ugs}, thus further
modifying the yield. How these loosely bound objects can be formed and survive in such a hostile environment is still an unsolved question~\cite{Andronic:2017pug}.
The coalescence model uses a different approach to explain the production of nuclei: the size of the nucleon-emitting source, accessible through the analysis of femtoscopic correlations~\cite{ALICE:2020ibs}, and the nuclear wave function are the two inputs that determine the formation probability of bound states~\cite{Blum:2017qnn,Sun:2018mqq}.
While the SHM can compute directly the absolute yields of particles, in the hadron coalescence model the yield of bound states can be computed only relative to the yields of other particles.

The measurement of the production of large bound states in small collision systems, such as pp and \pPb, is considered to allow for conclusive tests~\cite{Bellini:2018epz,Bellini:2020cbj} of  nucleosynthesis in hadronic collisions.
An extreme example is the hypertriton \hyp, the bound state of a proton, a neutron, and a $\Lambda$ baryon. This state is characterised by a very small $\Lambda$ separation energy, of the order of a few hundreds of keV~\cite{Davis:2005mb,Adam:2019phl}, and consequently it has a wide wave function that can extend up to a radius of $\approx$ 10 fm~\cite{Nemura:1999qp, Hildenbrand:2019sgp}.
The size of the \hyp wave function is therefore much larger than the hadron emission radius estimated with a femtoscopic technique in \pPb collisions (1--2 fm,~\cite{Abelev:2014pja,Adam:2015pya}). For this reason, the \hyp yield in \pPb collisions predicted by the coalescence model, where the ratio of nucleus size to source size directly influences its yield, is suppressed with respect to the statistical hadronisation model expectations, where the nuclear size does not enter explicitly~\cite{Vovchenko:2018fiy,Bellini:2018epz,Sun:2018mqq}.

The results presented in this Letter are based on data collected during the 2013 and 2016 \pPb LHC runs at \snn = 5.02 TeV. With this beam configuration, the nucleon--nucleon centre-of-mass system moves in rapidity by $\Delta y_{\text{cms}}$ = 0.465 in the direction of the proton beam. The ALICE detector and its performance are described in detail in ~\cite{Aamodt:2008zz,Abelev:2014ffa}.
Collision events are selected by using the information from the V0A and V0C scintillator arrays~\cite{Abbas:2013taa},  located on both sides of the interaction point, covering the pseudorapidity intervals
$-3.7 <\eta <-1.7$ and
$2.8<\eta<5.1$.
A coincident signal in both arrays is used as a minimum-bias (MB) trigger. In addition, only events with the primary vertex position within 10 cm along the beam axis to the nominal centre of the experiment are selected to benefit from the full acceptance of the detector.
Furthermore, to ensure the best possible performance of the detector and the proper normalisation of the results, events with more than one reconstructed primary interaction vertex (pile-up events) are rejected.
In total, about 750 million MB events are selected for analysis, corresponding to an integrated luminosity of $\mathcal{L}_{\rm{int}}^{\rm{MB}} = 359~
\mathrm{\mu b}^{-1}$, with a relative uncertainty determined by the van der
Meer scan to be 3.7\%~\cite{Abelev:2014epa}.
For this analysis, the 40\% of events with the highest multiplicity measured by the V0A detector are used.

The \hyp candidates are reconstructed via the charged two-body decay channel $\hyp \rightarrow \mathrm{^3He} + \pi^-$ (and the related charge conjugated particles for $\mathrm{^3_{\overline{\Lambda}} \overline{H}}$). In this work, \hyp and \antihyp are combined to reduce the statistical uncertainty.

\begin{figure}[!ht]
    \centering
    \includegraphics[width=0.49\textwidth]{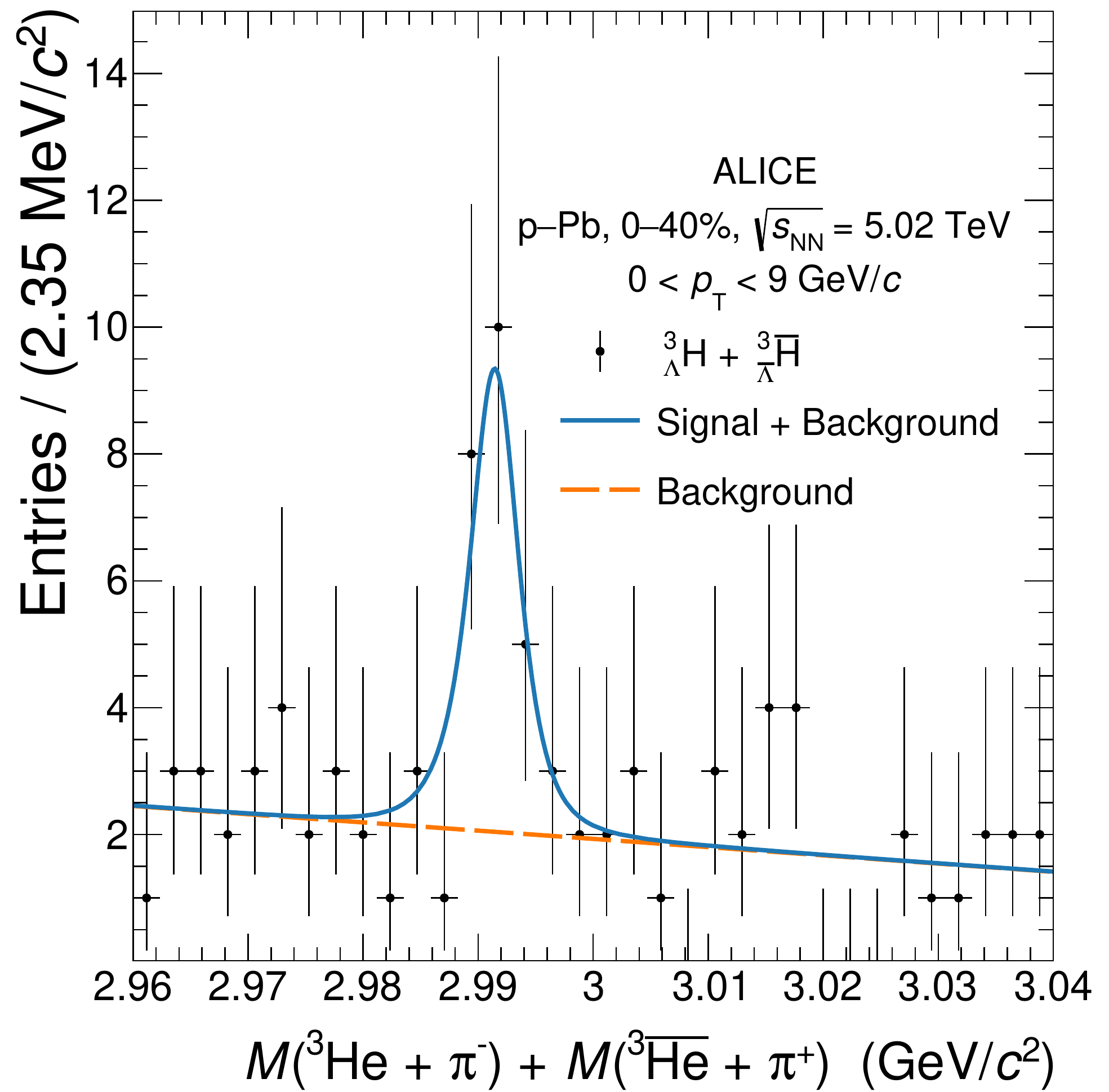}
    \caption{Invariant mass distribution of the $\Hee + \pi^{-}$ and charge conjugate pairs passing the analysis selections. Vertical lines represent the statistical Poissonian uncertainties. The invariant mass spectrum is fitted with a two-component model: the blue line represents the total fit while the orange dashed line shows the background component only.}
    \label{fig:sig_extr}
    \end{figure}

In the following, we use the notation \hyp and \Hee for both the particle and the antiparticle, as well as for their average.
The charged-particle tracks are reconstructed in the ALICE central barrel with the Inner Tracking System (ITS)~\cite{Aamodt:2010aa} and the Time Projection Chamber (TPC)~\cite{Alme:2010ke}, which are located within a solenoid that provides a homogeneous magnetic field of $0.5$ T in the direction of the beam axis.
These two subsystems provide full azimuthal coverage for charged-particle trajectories in the pseudorapidity interval $|\eta| < 0.8$.
The TPC is also used for the particle identification (PID) of the $^{3}$He and the $\pi^-$  via their specific energy loss \dEdx in the gas volume, with a \dEdx resolution of about $5\%$~\cite{Alme:2010ke}.
The $n(\sigma^{\rm TPC}_{i})$ variable represents the PID response in the TPC expressed in terms of the deviation between the measured and the expected \dEdx for a particle species $i$, normalized by the detector resolution $\sigma$. The expected \dEdx is computed with a parameterised Bethe-Bloch function~\cite{Abelev:2014ffa}. Pion and \Hee tracks within $5\sigma^{\rm TPC}$ are selected.
The identified \Hee and $\pi$ tracks are then used to reconstruct the \hyp weak decay topology with an algorithm similar to that used in previous analyses~\cite{Adam:2015yta, Acharya:2019qcp}. By combining the information on the decay kinematics and decay vertex, several selection variables are defined. Those used in the analysis are: the radial distance of the decay vertex from the beam line, the distance of each daughter track from both the primary and the decay vertex, the proper decay length of the candidate ($ct$) and cos($\mathrm{\theta_P}$), where $\mathrm{\theta_P}$ is the angle between the total momentum vector of the decay daughters and the straight line connecting the primary and secondary vertices.
The final candidate selection based on these variables is performed with a gradient boosted decision tree classifier (BDT) implemented by the XGBoost library~\cite{10.1145/2939672.2939785,ATLAS:2018mme, barioglio_luca_2021_5734093} and trained on a dedicated Monte Carlo (MC) simulated event sample. The MC sample is created using the HIJING event generator~\cite{Wang:1991hta} for simulating the underlying \pPb collisions, while \hyp were injected with a \pt distribution represented by a $m\mathrm{_{T}}$ exponential function that describes the \pt distribution of \Hee as measured in \pPb collisions~\cite{Acharya:2019xmu}. The particles are transported through the detector geometry using \textsc{Geant4}~\cite{Agostinelli:2002hh}, which simulates the interaction with the material and the weak decay of the \hyp.
 The BDT is a supervised learning algorithm that determines how to
discriminate between two or more classes, signal and background in this case, by
examining sets of examples called the training sets. In this analysis, the training sets are composed of \hyp signal candidates extracted from the MC sample and background candidates from paired like-sign $\mathrm{^3He}$ and $\mathrm{\pi}$ tracks from the data.
For each \hyp candidate, the BDT combines topological and single track variables to return a score, which is proportional to the candidate probability of being signal or background.
The selection is based on the BDT score, defining a threshold that maximises the expected signal significance assuming thermal production. In this analysis the default BDT score selection corresponds to a 72\% signal efficiency and a $\mathrm{{3} x 10^{-5}}$ background rejection factor.
The candidates that pass the BDT selection are used to populate the invariant mass distribution in the transverse momentum interval 0 $<$ \pt $<$ 9 GeV/$c$. An excess of entries is observed at a mass near 2.99 GeV/$c^2$, as shown in Fig.~\ref{fig:sig_extr}. The unbinned distribution is fitted with a Kernel Density Estimator (KDE)~\cite{Cranmer:2000du, Verkerke:2003ir} function tuned on the MC sample to describe the signal and a linear function to describe the background component. The KDE is chosen for smoothing the template extracted from the MC.
The invariant mass distribution with the superimposed fit is shown in Fig.~\ref{fig:sig_extr}.

The significance associated with the signal is evaluated following the procedure described in~\cite{Cowan:2010js}: the probability for a background fluctuation to be at least as large as the observed maximum excess (local p-value) is computed by employing the asymptotic formulae for likelihood-based tests. The local p-value is expressed as a corresponding number of standard deviations using the one-sided Gaussian tail convention. The excess  of entries observed above the expected background has a local significance of 4.6 standard deviations at the nominal \hyp mass.
The production yield is obtained starting from the signal extracted from the fit to the invariant mass spectrum. Then the fitted signal is corrected for the reconstruction and the selection efficiency, including reconstruction efficiencies for the daughter particles and the topology, the acceptance of the ALICE detector, the number of analysed events, the branching ratio (B.R.) of the \hyp in the two-body decay channel and the fraction of \hyp that are absorbed in the ALICE detector ($f_{\rm abs}$). The simulation of inelastic interactions of the daughter particles is done with \textsc{Geant4} and is taken into account in the reconstruction efficiency computation. The B.R. value is assumed to be 0.25 according to the calculation published in~\cite{Kamada:1997rv}.

The systematic uncertainties originate from (1) the \hyp selection and the signal extraction, (2) the choice of the \hyp input \pt distribution in the Monte Carlo sample, and (3) the \hyp absorption in the detector. In addition (4), a $9\%$ systematic uncertainty is added due to the uncertainty of the B.R. as explained later in the text. The total uncertainty is obtained as the quadratic sum of the individual contributions. The first contribution, which is the dominant one, is computed by varying simultaneously the BDT threshold ($\pm 5\%$) and the background fit function (constant, linear, exponential). The standard deviation (RMS) of the different yields represents our systematic error associated with the BDT selection and the signal extraction, and it amounts to 14\%. The second contribution is evaluated
by using different input \pt distributions for the Monte Carlo sample and evaluating the
effects on the efficiency. Four different \pt models ($\mathrm{m_T}$ exponential, \pt exponential, Boltzmann and Blast Wave~\cite{Schnedermann:1993ws}) are fitted to the \he \pt distribution~\cite{Acharya:2019xmu}. For each of them the efficiency and the yield are computed assuming that the \he and the \hyp have the same \pt distributions as already seen for light flavour hadrons with similar masses in all collision systems~\cite{Adam:2015vda,Adam:2015yta,Abelev:2013haa}. The RMS among the trials is calculated, yielding a systematic uncertainty of $7\%$. Finally, the uncertainty of $f_{\rm abs}$ is considered. According to~\cite{Evlanov:1998py}, the expected absorption cross section of \hyp due to the inelastic interactions in the ALICE detector material is ${\approx}$ 1.5 times that of \antiHee ($\mathrm{\sigma_{inel}^{\antiHee}}$). The value of $f_{\rm abs}$ is computed by simulating the passage of hypertritons through the ALICE detector using this cross section and gives a result of $\approx$ 3$\%$. The systematic uncertainty of $f_{\rm abs}$ is evaluated by employing different cross sections for the \hyp from zero (no interactions) to $2\mathrm{\sigma_{inel}^{\Hee}}$. For each variation $f_{\rm abs}$ is recalculated. This results in a systematic uncertainty on the yield of about $4 \%$. Larger variations of the inelastic cross section are here not considered as they spoil the exponential trend of the proper decay length spectrum measured in Pb--Pb collisions.

The resulting corrected \hyp yield in the rapidity interval $-1<y<0$ together with its statistical and systematic uncertainties is
\[\frac{\dd N}{\dd y} = [\mathrm{6.3 \pm 1.8 (stat.) \pm 1.2 (syst.) ] x 10^{-7}}.\]

\begin{figure*}[!t]
    \centering
    \includegraphics[width=.49\textwidth]{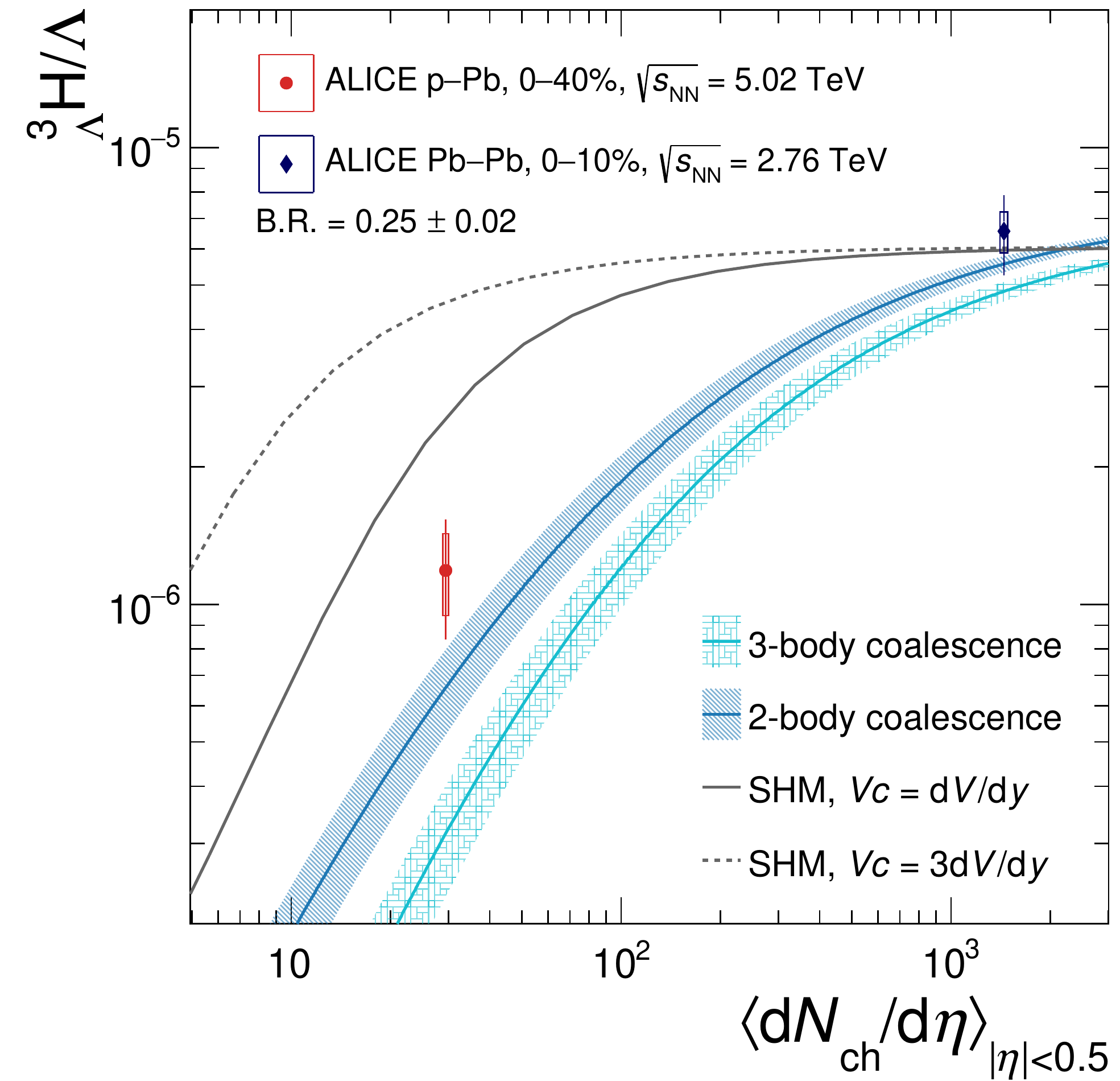}
    \includegraphics[width=.49\textwidth]{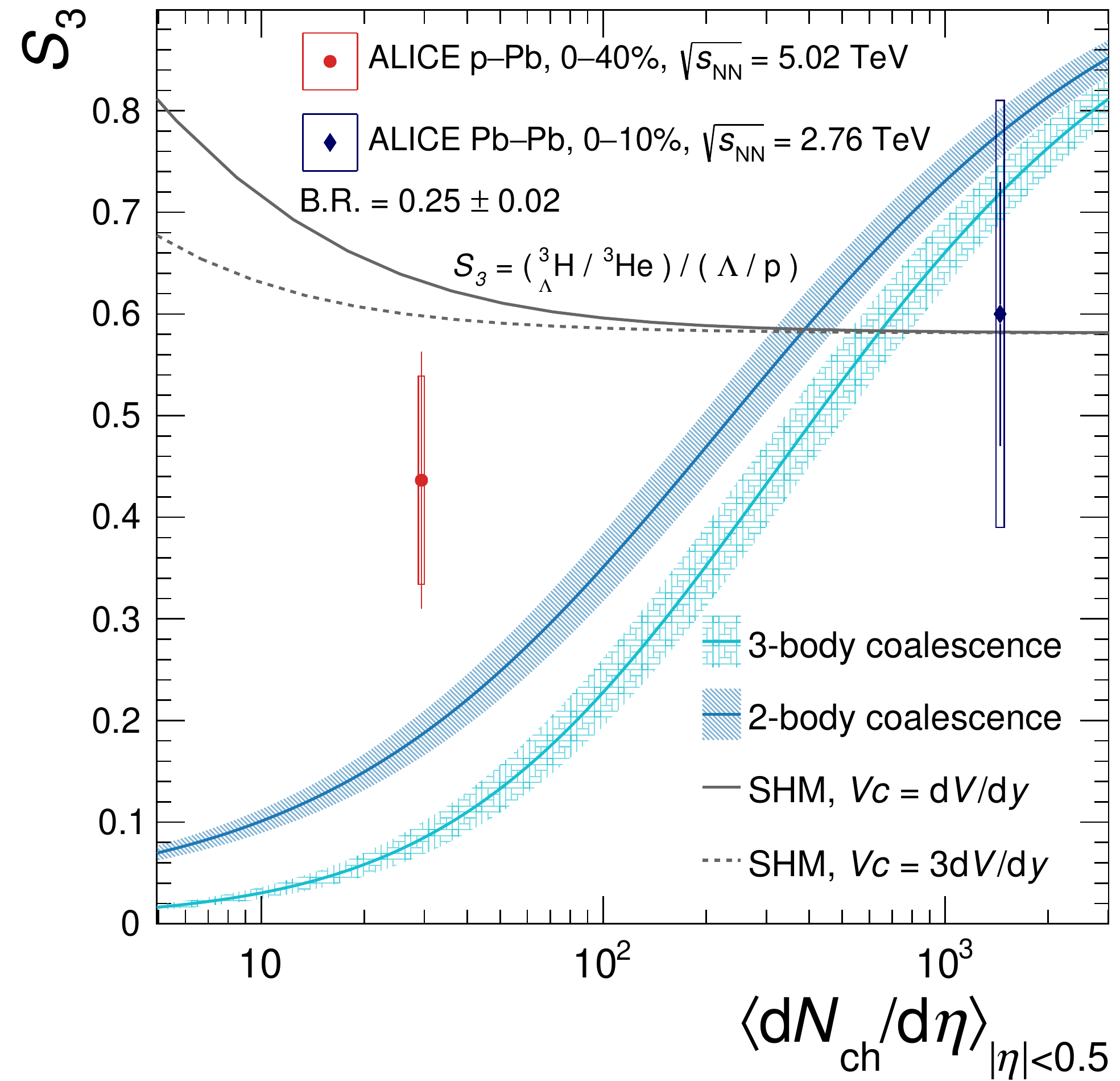}
    \caption{\hyp/$\Lambda$ (on the left) and $S_3$ (on the right) measurements in p--Pb (in red) and Pb--Pb collisions~\cite{Adam:2015yta} (in blue) as a function of mean charged-particle multiplicity. The vertical lines and boxes are the statistical and systematic uncertainties (including the uncertainty on the B.R.), respectively.  The expectations for the canonical statistical hadronization~\cite{Vovchenko:2018fiy} and coalescence models are shown~\cite{Sun:2018mqq}.}
    \label{fig:s3}
\end{figure*}

The result is compared with the expectations from the canonical SHM~\cite{Vovchenko:2018fiy}, which assumes exact conservation of baryon number, strangeness, and electric charge across a correlation volume $V_{c}$. The SHM predictions are computed using a fixed chemical freeze-out temperature of $T_{\rm chem}$ = 155 MeV and two correlation volumes extending across one unit ($V_{c} = \dd V/\dd y$), and three units ($V_{c} = 3\dd V/\dd y$) of rapidity~\cite{Vovchenko:2018fiy}. The size of the correlation volume governs the influence of exact quantum number conservation, with smaller values leading to a stronger suppression of conserved charges and $V_{c} \rightarrow \infty$ leading to the grand canonical ensemble.
The \hyp \pt integrated yield is $\mathrm{{1.1} x 10^{-6}}$ and $\mathrm{2.0 x 10^{-6}}$ with $V_{c} = \dd V/\dd y$ and $V_{c} = 3\dd V/\dd y$, respectively.
The $\dd N / \dd y$ predictions by the model were obtained using the code released together with the publication~\cite{Vovchenko:2019pjl}.

As explained above, in the case of the coalescence model it is not possible to compare directly the measured absolute yield to the model prediction.
Hence, this comparison is attained by computing the \hyp$/\Lambda$ ratio and the strangeness population factor $S_3 = (\hyp/\Hee)$/$(\mathrm{\Lambda/p})$~\cite{Zhang:2009ba} using previous ALICE measurements of p, $\Lambda$, and \Hee yields~\cite{Acharya:2019xmu, Abelev:2013haa}, as shown in Fig.~\ref{fig:s3}. The yield of the $\Lambda$ baryon, measured in $-0.5<y<0$, has been extrapolated to the \hyp rapidity region using MC generators~\cite{Gyulassy:1994ew,Pierog:2013ria,Roesler:2000he} that are known to reproduce the pseudorapidity density distribution of charged hadrons~\cite{ALICE:2012xs}. The corresponding correction is approximately 2\%.
In central \PbPb collisions the data are consistent with both coalescence and SHM predictions, which are similar, as shown in Fig.~\ref{fig:s3}.  The situation is different for \pPb collisions where the two models are well separated. Taking into account the uncertainties of the measurement as well as the model uncertainty, the measured $S_3$ ratio is compatible with the two-body (deuteron-$\Lambda$) and three-body (proton-neutron-$\Lambda$) coalescence within 1.2$\sigma$ and 2$\sigma$, respectively.
With its large uncertainties, also due to the large uncertainty on the \Hee yield, the $S_3$ is compatible within 2$\sigma$ with the SHM calculations too. Hence, the \hyp$/\Lambda$ ratio is used as a test for coalescence and SHM predictions in the following. In this case, the measurement is deviating by 3.2$\sigma$ and 7.9$\sigma$ from the SHM with  $V_{c}=1\dd V/\dd y$ and $V_{c}=3\dd V/\dd y$, respectively. On the other hand, both the coalescence calculations are within 2$\sigma$ of the measured \hyp$/\Lambda$.
It has to be noted that recent measurements of the \hyp mass~\cite{Adam:2019phl} suggest a larger binding energy, and hence a smaller wave function, of the \hyp. This would further shift upward the coalescence predictions.

\begin{figure}[!ht]
    \centering
    \includegraphics[width=.49\textwidth]{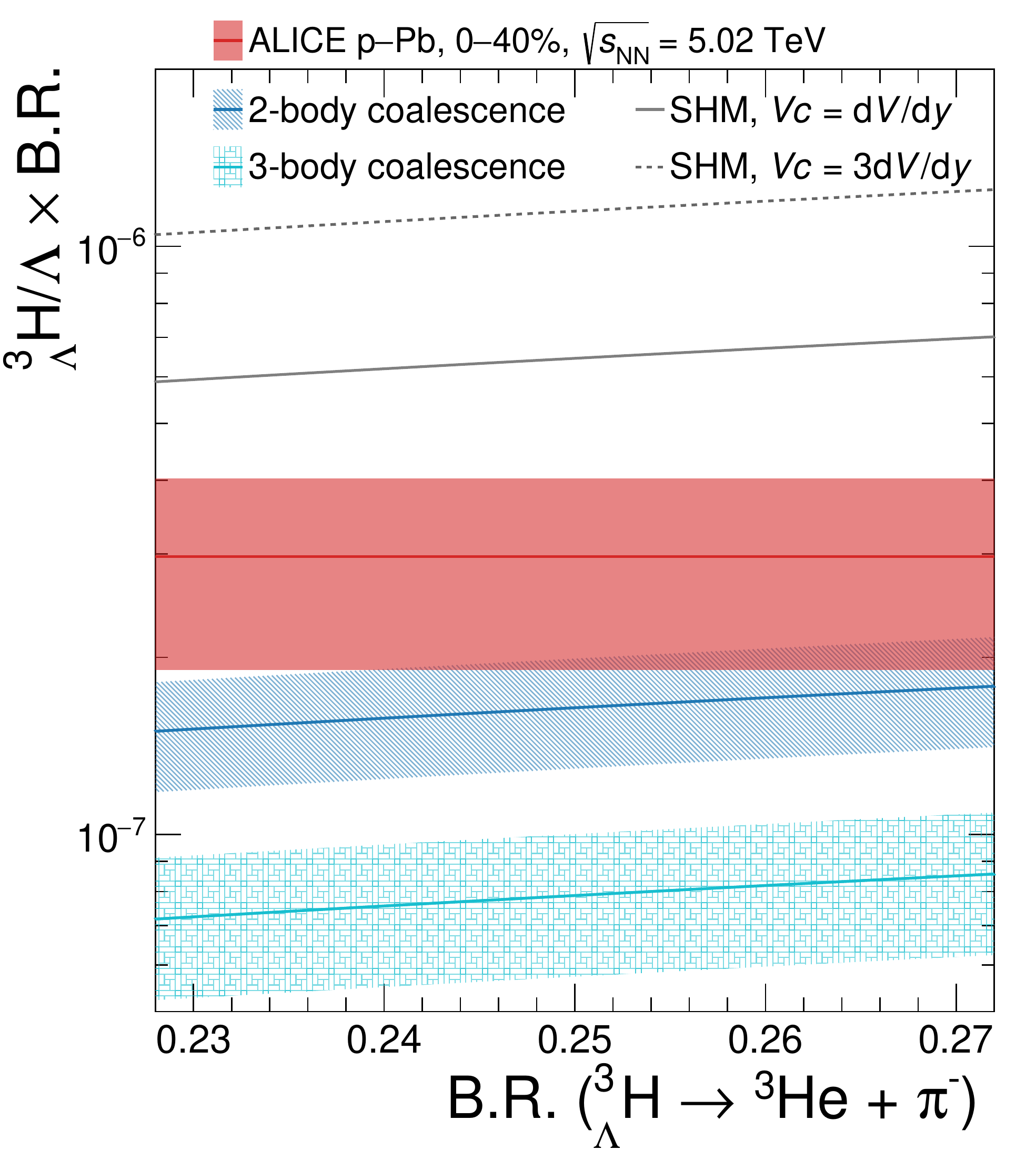}
    \caption{\hyp/$\Lambda$ times branching ratio as a function of branching ratio. The horizontal line is the measured value and the band represents statistical and systematic uncertainties added in quadrature. The expectations for the canonical statistical hadronization~\cite{Vovchenko:2018fiy} and coalescence models are shown~\cite{Sun:2018mqq}.}
    \label{fig:br_ratio}
\end{figure}

The value of  B.R. = 0.25 for the \hyp$\rightarrow \mathrm{^3He}+\pi$ decay used in this analysis was computed theoretically in Ref.~\cite{Kamada:1997rv}.
To investigate the uncertainty resulting from this assumption, Fig.~\ref{fig:br_ratio} shows the measured \hyp/$\Lambda \,\times\,\text{B.R.}$ for different theoretical model calculations~\cite{Vovchenko:2018fiy,Sun:2018mqq} assuming a possible variation of the B.R. value. The variation range is chosen by evaluating the relative deviation between the theoretical $\mathrm{R_{3}}$  and the world average of all the $\mathrm{R_{3}}$ measurements including the most recent measurement in heavy-ion collisions~\cite{Adamczyk:2017buv}, where $\mathrm{R_{3}}$ is defined as:
\[\mathrm{R_{3} = \frac{
\Gamma(\hyp \rightarrow \mathrm{^3He} + \pi^-)}{\Gamma(\hyp \rightarrow all \ \pi^- \ decay \ channels)}}.\]
This uncertainty on $\mathrm{R_3}$ is propagated to the $\mathrm{B.R.(\hyp \rightarrow \mathrm{^3He} + \pi^-)}$ and corresponds to a variation range of $\pm 9\%$ around the nominal value.
While the two-body coalescence calculation is compatible with the data for the nominal or larger B.R., a discrepancy of 2$\sigma$ is observed between data and the three-body coalescence prediction. Furthermore, in the whole B.R. variation interval, the SHM is more than 2.7$\sigma$ and 6.9$\sigma$ away from the measured \hyp/$\Lambda\,\times\,\text{B.R.}$ for the $V_c=1\dVdy$ and $V_c=3\dVdy$ configurations, respectively.

In summary, the first measurement of the production yield of hypertriton in p--Pb collisions at the LHC is reported. The measurements of yields of \hyp in \pPb collisions provide an
opportunity to potentially discriminate between nucleosynthesis models. The measured \pt integrated yield excludes, with high significance, canonical versions of the SHM with $V_{c}\geq3\dd V/\dd y$ to explain the (hyper)nuclei production in \pPb collisions. It remains to be seen if advanced versions of the SHM using the S-matrix approach to account for the interactions among hadrons~\cite{Cleymans:2020fsc} will be able to solve this discrepancy. The $\hyp/\Lambda$ ratio is well described by the two-body coalescence prediction, while the three-body formulation is slightly disfavoured by our measurement.
While the general conclusions of the comparison with the models are unaltered even when considering large variations of the $\mathrm{B.R.(\hyp \rightarrow \mathrm{^3He} + \pi^-)}$ around the value available in literature, the significance of the comparison between data and models is influenced by this uncertainty. Upcoming studies using the LHC Run 2 Pb--Pb data will help to reduce this uncertainty by measuring the $\hyp\rightarrow\mathrm{d+p+\pi^-}$ decay channel relative branching ratio.
Furthermore, with the upgraded ALICE apparatus and the upcoming LHC Run 3, it will be possible to reduce both the statistical and the systematic uncertainties of the \hyp yield measurements in pp~\cite{ALICE-PUBLIC-2020-005} and \pPb collisions and to study the \hyp production as a function of the size of the nucleon-emitting source measured with femtoscopic correlations. These studies may make it possible to decisively distinguish between the two production models.

%% file: fa_2021-07-06.tex

The ALICE Collaboration would like to thank all its engineers and technicians for their invaluable contributions to the construction of the experiment and the CERN accelerator teams for the outstanding performance of the LHC complex.
The ALICE Collaboration gratefully acknowledges the resources and support provided by all Grid centres and the Worldwide LHC Computing Grid (WLCG) collaboration.
The ALICE Collaboration acknowledges the following funding agencies for their support in building and running the ALICE detector:
A. I. Alikhanyan National Science Laboratory (Yerevan Physics Institute) Foundation (ANSL), State Committee of Science and World Federation of Scientists (WFS), Armenia;
Austrian Academy of Sciences, Austrian Science Fund (FWF): [M 2467-N36] and Nationalstiftung f\"{u}r Forschung, Technologie und Entwicklung, Austria;
Ministry of Communications and High Technologies, National Nuclear Research Center, Azerbaijan;
Conselho Nacional de Desenvolvimento Cient\'{\i}fico e Tecnol\'{o}gico (CNPq), Financiadora de Estudos e Projetos (Finep), Funda\c{c}\~{a}o de Amparo \`{a} Pesquisa do Estado de S\~{a}o Paulo (FAPESP) and Universidade Federal do Rio Grande do Sul (UFRGS), Brazil;
Ministry of Education of China (MOEC) , Ministry of Science \& Technology of China (MSTC) and National Natural Science Foundation of China (NSFC), China;
Ministry of Science and Education and Croatian Science Foundation, Croatia;
Centro de Aplicaciones Tecnol\'{o}gicas y Desarrollo Nuclear (CEADEN), Cubaenerg\'{\i}a, Cuba;
Ministry of Education, Youth and Sports of the Czech Republic, Czech Republic;
The Danish Council for Independent Research | Natural Sciences, the VILLUM FONDEN and Danish National Research Foundation (DNRF), Denmark;
Helsinki Institute of Physics (HIP), Finland;
Commissariat \`{a} l'Energie Atomique (CEA) and Institut National de Physique Nucl\'{e}aire et de Physique des Particules (IN2P3) and Centre National de la Recherche Scientifique (CNRS), France;
Bundesministerium f\"{u}r Bildung und Forschung (BMBF) and GSI Helmholtzzentrum f\"{u}r Schwerionenforschung GmbH, Germany;
General Secretariat for Research and Technology, Ministry of Education, Research and Religions, Greece;
National Research, Development and Innovation Office, Hungary;
Department of Atomic Energy Government of India (DAE), Department of Science and Technology, Government of India (DST), University Grants Commission, Government of India (UGC) and Council of Scientific and Industrial Research (CSIR), India;
Indonesian Institute of Science, Indonesia;
Istituto Nazionale di Fisica Nucleare (INFN), Italy;
Institute for Innovative Science and Technology , Nagasaki Institute of Applied Science (IIST), Japanese Ministry of Education, Culture, Sports, Science and Technology (MEXT) and Japan Society for the Promotion of Science (JSPS) KAKENHI, Japan;
Consejo Nacional de Ciencia (CONACYT) y Tecnolog\'{i}a, through Fondo de Cooperaci\'{o}n Internacional en Ciencia y Tecnolog\'{i}a (FONCICYT) and Direcci\'{o}n General de Asuntos del Personal Academico (DGAPA), Mexico;
Nederlandse Organisatie voor Wetenschappelijk Onderzoek (NWO), Netherlands;
The Research Council of Norway, Norway;
Commission on Science and Technology for Sustainable Development in the South (COMSATS), Pakistan;
Pontificia Universidad Cat\'{o}lica del Per\'{u}, Peru;
Ministry of Education and Science, National Science Centre and WUT ID-UB, Poland;
Korea Institute of Science and Technology Information and National Research Foundation of Korea (NRF), Republic of Korea;
Ministry of Education and Scientific Research, Institute of Atomic Physics and Ministry of Research and Innovation and Institute of Atomic Physics, Romania;
Joint Institute for Nuclear Research (JINR), Ministry of Education and Science of the Russian Federation, National Research Centre Kurchatov Institute, Russian Science Foundation and Russian Foundation for Basic Research, Russia;
Ministry of Education, Science, Research and Sport of the Slovak Republic, Slovakia;
National Research Foundation of South Africa, South Africa;
Swedish Research Council (VR) and Knut \& Alice Wallenberg Foundation (KAW), Sweden;
European Organization for Nuclear Research, Switzerland;
Suranaree University of Technology (SUT), National Science and Technology Development Agency (NSDTA) and Office of the Higher Education Commission under NRU project of Thailand, Thailand;
Turkish Energy, Nuclear and Mineral Research Agency (TENMAK), Turkey;
National Academy of  Sciences of Ukraine, Ukraine;
Science and Technology Facilities Council (STFC), United Kingdom;
National Science Foundation of the United States of America (NSF) and United States Department of Energy, Office of Nuclear Physics (DOE NP), United States of America.

%% file: Alice_Authorlist_2021-07-06.tex
\begin{flushleft}

S.~Acharya$^{\rm 143}$, 
D.~Adamov\'{a}$^{\rm 98}$, 
A.~Adler$^{\rm 76}$, 
G.~Aglieri Rinella$^{\rm 35}$, 
M.~Agnello$^{\rm 31}$, 
N.~Agrawal$^{\rm 55}$, 
Z.~Ahammed$^{\rm 143}$, 
S.~Ahmad$^{\rm 16}$, 
S.U.~Ahn$^{\rm 78}$, 
I.~Ahuja$^{\rm 39}$
% ...

\bigskip 

% ...
$^{16}$ Department of Physics, Aligarh Muslim University, Aligarh, India\\
% ...
$^{31}$ Dipartimento DISAT del Politecnico and Sezione INFN, Turin, Italy\\
% ...
$^{35}$ European Organization for Nuclear Research (CERN), Geneva, Switzerland\\
% ...
$^{39}$ Faculty of Science, P.J.~\v{S}af\'{a}rik University, Ko\v{s}ice, Slovakia\\
% ...
$^{55}$ INFN, Sezione di Bologna, Bologna, Italy\\
% ...
$^{76}$ Johann-Wolfgang-Goethe Universit\"{a}t Frankfurt Institut f\"{u}r Informatik, Fachbereich Informatik und Mathematik, Frankfurt, Germany\\
% ...
$^{78}$ Korea Institute of Science and Technology Information, Daejeon, Republic of Korea\\
% ...
$^{98}$ Nuclear Physics Institute of the Czech Academy of Sciences, \v{R}e\v{z} u Prahy, Czech Republic\\
% ...
$^{143}$ Variable Energy Cyclotron Centre, Homi Bhabha National Institute, Kolkata, India\\
% ...

\end{flushleft} 

%% file: main.bbl
\providecommand{\href}[2]{#2}\begingroup\raggedright\begin{thebibliography}{10}

\bibitem{Adam:2015vda}
{\bfseries ALICE} Collaboration, J.~Adam {\em et~al.}, ``{Production of light
  nuclei and anti-nuclei in pp and Pb--Pb collisions at energies available at
  the CERN Large Hadron Collider}'',
  \href{http://dx.doi.org/10.1103/PhysRevC.93.024917}{{\em Phys. Rev.}
  {\bfseries C93} (2016) 024917},
\href{http://arxiv.org/abs/1506.08951}{{\ttfamily arXiv:1506.08951 [nucl-ex]}}.

\bibitem{Acharya:2017dmc}
{\bfseries ALICE} Collaboration, S.~Acharya {\em et~al.}, ``{Measurement of
  deuteron spectra and elliptic flow in Pb–Pb collisions at $\sqrt{s_{\mathrm
  {NN}}}$ = 2.76 TeV at the LHC}'',
  \href{http://dx.doi.org/10.1140/epjc/s10052-017-5222-x}{{\em Eur. Phys. J.}
  {\bfseries C77} (2017) 658},
\href{http://arxiv.org/abs/1707.07304}{{\ttfamily arXiv:1707.07304 [nucl-ex]}}.

\bibitem{Acharya:2019rgc}
{\bfseries ALICE} Collaboration, S.~Acharya {\em et~al.}, ``{Multiplicity
  dependence of (anti-)deuteron production in pp collisions at $\sqrt{s}$ = 7
  TeV}'', \href{http://dx.doi.org/10.1016/j.physletb.2019.05.028}{{\em Phys.
  Lett.} {\bfseries B794} (2019) 50--63},
\href{http://arxiv.org/abs/1902.09290}{{\ttfamily arXiv:1902.09290 [nucl-ex]}}.

\bibitem{Acharya:2019rys}
{\bfseries ALICE} Collaboration, S.~Acharya {\em et~al.}, ``{Multiplicity
  dependence of light (anti-)nuclei production in p-Pb collisions at
  $\sqrt{s_{\rm{NN}}}$ = 5.02 TeV}'',
  \href{http://dx.doi.org/10.1016/j.physletb.2019.135043}{{\em Phys. Lett.}
  {\bfseries B800} (2020) 135043},
\href{http://arxiv.org/abs/1906.03136}{{\ttfamily arXiv:1906.03136 [nucl-ex]}}.

\bibitem{Acharya:2019xmu}
{\bfseries ALICE} Collaboration, S.~Acharya {\em et~al.}, ``{Production of
  (anti-)$^3$He and (anti-)$^3$H in p-Pb collisions at $\sqrt{s_{\rm{NN}}}$ =
  5.02 TeV}'', \href{http://dx.doi.org/10.1103/PhysRevC.101.044906}{{\em Phys.
  Rev. C} {\bfseries 101} (2020) 044906},
  \href{http://arxiv.org/abs/1910.14401}{{\ttfamily arXiv:1910.14401
  [nucl-ex]}}.

\bibitem{Acharya:2020sfy}
{\bfseries ALICE} Collaboration, S.~Acharya {\em et~al.}, ``{(Anti-)deuteron
  production in pp collisions at $\sqrt{s}=13 \ \text {TeV}$}'',
  \href{http://dx.doi.org/10.1140/epjc/s10052-020-8256-4}{{\em Eur. Phys. J. C}
  {\bfseries 80} (2020) 889}, \href{http://arxiv.org/abs/2003.03184}{{\ttfamily
  arXiv:2003.03184 [nucl-ex]}}.

\bibitem{Acharya:2017bso}
{\bfseries ALICE} Collaboration, S.~Acharya {\em et~al.}, ``{Production of
  $^{4}$He and $^{4}\overline{\textrm{He}}$ in Pb-Pb collisions at
  $\sqrt{s_{\mathrm{NN}}}$ = 2.76 TeV at the LHC}'',
  \href{http://dx.doi.org/10.1016/j.nuclphysa.2017.12.004}{{\em Nucl. Phys. A}
  {\bfseries 971} (2018) 1--20},
  \href{http://arxiv.org/abs/1710.07531}{{\ttfamily arXiv:1710.07531
  [nucl-ex]}}.

\bibitem{Schael:2006fd}
{\bfseries ALEPH} Collaboration, S.~Schael {\em et~al.}, ``{Deuteron and
  anti-deuteron production in e+ e- collisions at the Z resonance}'',
  \href{http://dx.doi.org/10.1016/j.physletb.2006.06.043}{{\em Phys. Lett. B}
  {\bfseries 639} (2006) 192--201},
  \href{http://arxiv.org/abs/hep-ex/0604023}{{\ttfamily arXiv:hep-ex/0604023}}.

\bibitem{Alper:1973my}
{\bfseries British-Scandinavian} Collaboration, B.~Alper {\em et~al.}, ``{Large
  angle production of stable particles heavier than the proton and a search for
  quarks at the cern intersecting storage rings}'',
\href{http://dx.doi.org/10.1016/0370-2693(73)90700-4}{{\em Phys. Lett.}
  {\bfseries 46B} (1973) 265--268}.

\bibitem{Henning:1977mt}
{\bfseries British-Scandinavian-MIT} Collaboration, S.~Henning {\em et~al.},
  ``{Production of Deuterons and anti-Deuterons in Proton Proton Collisions at
  the CERN ISR}'',
\href{http://dx.doi.org/10.1007/BF02822248}{{\em Lett. Nuovo Cim.} {\bfseries
  21} (1978) 189}.

\bibitem{Alexopoulos:2000jk}
{\bfseries E735} Collaboration, T.~Alexopoulos {\em et~al.}, ``{Cross-sections
  for deuterium, tritium, and helium production in p p collisions at s =
  1.8-TeV}'', \href{http://dx.doi.org/10.1103/PhysRevD.62.072004}{{\em Phys.
  Rev. D} {\bfseries 62} (2000) 072004}.

\bibitem{Adler:2001uy}
{\bfseries STAR} Collaboration, C.~Adler {\em et~al.}, ``{Anti-deuteron and
  anti-{$^3$He} production in $\snn~=~130$~GeV Au+Au collisions}'',
  \href{http://dx.doi.org/10.1103/PhysRevLett.87.262301}{{\em Phys. Rev. Lett.}
  {\bfseries 87} (2001) 262301},
  \href{http://arxiv.org/abs/nucl-ex/0108022}{{\ttfamily arXiv:nucl-ex/0108022
  [nucl-ex]}}.
[Erratum: Phys. Rev. Lett.87,279902(2001)].

\bibitem{Adler:2004uy}
{\bfseries PHENIX} Collaboration, S.~S. Adler {\em et~al.}, ``{Deuteron and
  antideuteron production in Au + Au collisions at $\snn~=~200$~GeV}'',
  \href{http://dx.doi.org/10.1103/PhysRevLett.94.122302}{{\em Phys. Rev. Lett.}
  {\bfseries 94} (2005) 122302},
\href{http://arxiv.org/abs/nucl-ex/0406004}{{\ttfamily arXiv:nucl-ex/0406004
  [nucl-ex]}}.

\bibitem{Aktas:2004pq}
{\bfseries H1} Collaboration, A.~Aktas {\em et~al.}, ``{Measurement of
  anti-deuteron photoproduction and a search for heavy stable charged particles
  at HERA}'', \href{http://dx.doi.org/10.1140/epjc/s2004-01978-x}{{\em Eur.
  Phys. J. C} {\bfseries 36} (2004) 413--423},
  \href{http://arxiv.org/abs/hep-ex/0403056}{{\ttfamily arXiv:hep-ex/0403056}}.

\bibitem{Asner:2006pw}
{\bfseries CLEO} Collaboration, D.~M. Asner {\em et~al.}, ``{Anti-deuteron
  production in Upsilon(nS) decays and the nearby continuum}'',
  \href{http://dx.doi.org/10.1103/PhysRevD.75.012009}{{\em Phys. Rev. D}
  {\bfseries 75} (2007) 012009},
  \href{http://arxiv.org/abs/hep-ex/0612019}{{\ttfamily arXiv:hep-ex/0612019}}.

\bibitem{Agakishiev:2011ib}
{\bfseries STAR} Collaboration, H.~Agakishiev {\em et~al.}, ``{Observation of
  the antimatter helium-4 nucleus}'',
  \href{http://dx.doi.org/10.1038/nature10079}{{\em Nature} {\bfseries 473}
  (2011) 353}, \href{http://arxiv.org/abs/1103.3312}{{\ttfamily arXiv:1103.3312
  [nucl-ex]}}. [Erratum: Nature 475, 412 (2011)].

\bibitem{SimonGillo:1995dh}
{\bfseries NA44} Collaboration, J.~Simon-Gillo {\em et~al.}, ``{Deuteron and
  anti-deuteron production in CERN experiment NA44}'',
\href{http://dx.doi.org/10.1016/0375-9474(95)00259-4}{{\em Nucl. Phys.}
  {\bfseries A590} (1995) 483C--486C}.

\bibitem{Armstrong:2000gd}
{\bfseries E864} Collaboration, T.~A. Armstrong {\em et~al.}, ``{Anti-deuteron
  yield at the AGS and coalescence implications}'',
  \href{http://dx.doi.org/10.1103/PhysRevLett.85.2685}{{\em Phys. Rev. Lett.}
  {\bfseries 85} (2000) 2685--2688},
\href{http://arxiv.org/abs/nucl-ex/0005001}{{\ttfamily arXiv:nucl-ex/0005001
  [nucl-ex]}}.

\bibitem{Afanasev:2000ku}
{\bfseries NA49} Collaboration, S.~V. Afanasev {\em et~al.}, ``{Deuteron
  production in central Pb + Pb collisions at 158-A-GeV}'',
\href{http://dx.doi.org/10.1016/S0370-2693(00)00746-2}{{\em Phys. Lett.}
  {\bfseries B486} (2000) 22--28}.

\bibitem{Anticic:2004yj}
{\bfseries NA49} Collaboration, T.~Anticic {\em et~al.}, ``{Energy and
  centrality dependence of deuteron and proton production in Pb + Pb collisions
  at relativistic energies}'',
\href{http://dx.doi.org/10.1103/PhysRevC.69.024902}{{\em Phys. Rev.} {\bfseries
  C69} (2004) 024902}.

\bibitem{Mrowczynski:1987oid}
S.~Mrowczynski, ``{Deuteron formation mechanism}'',
  \href{http://dx.doi.org/10.1088/0305-4616/13/9/011}{{\em J. Phys. G}
  {\bfseries 13} (1987) 1089--1097}.

\bibitem{Scheibl:1998tk}
R.~Scheibl and U.~W. Heinz, ``{Coalescence and flow in ultrarelativistic heavy
  ion collisions}'', \href{http://dx.doi.org/10.1103/PhysRevC.59.1585}{{\em
  Phys. Rev.} {\bfseries C59} (1999) 1585--1602},
\href{http://arxiv.org/abs/nucl-th/9809092}{{\ttfamily arXiv:nucl-th/9809092
  [nucl-th]}}.

\bibitem{Sun:2018mqq}
K.-J. Sun, C.~M. Ko, and B.~D\"onigus, ``{Suppression of light nuclei
  production in collisions of small systems at the Large Hadron Collider}'',
  \href{http://dx.doi.org/10.1016/j.physletb.2019.03.033}{{\em Phys. Lett. B}
  {\bfseries 792} (2019) 132--137},
  \href{http://arxiv.org/abs/1812.05175}{{\ttfamily arXiv:1812.05175
  [nucl-th]}}.

\bibitem{Andronic:2010qu}
A.~Andronic, P.~Braun-Munzinger, J.~Stachel, and H.~Stocker, ``{Production of
  light nuclei, hypernuclei and their antiparticles in relativistic nuclear
  collisions}'', \href{http://dx.doi.org/10.1016/j.physletb.2011.01.053}{{\em
  Phys. Lett. B} {\bfseries 697} (2011) 203--207},
  \href{http://arxiv.org/abs/1010.2995}{{\ttfamily arXiv:1010.2995 [nucl-th]}}.

\bibitem{Vovchenko:2018fiy}
V.~Vovchenko, B.~D{\"o}nigus, and H.~Stoecker, ``{Multiplicity dependence of
  light nuclei production at LHC energies in the canonical statistical
  model}'', \href{http://dx.doi.org/10.1016/j.physletb.2018.08.041}{{\em Phys.
  Lett.} {\bfseries B785} (2018) 171--174},
\href{http://arxiv.org/abs/1808.05245}{{\ttfamily arXiv:1808.05245 [hep-ph]}}.

\bibitem{Blum:2017qnn}
K.~Blum, K.~C.~Y. Ng, R.~Sato, and M.~Takimoto, ``{Cosmic rays, antihelium, and
  an old navy spotlight}'',
  \href{http://dx.doi.org/10.1103/PhysRevD.96.103021}{{\em Phys. Rev.}
  {\bfseries D96} (2017) 103021},
\href{http://arxiv.org/abs/1704.05431}{{\ttfamily arXiv:1704.05431
  [astro-ph.HE]}}.

\bibitem{Korsmeier:2017xzj}
M.~Korsmeier, F.~Donato, and N.~Fornengo, ``{Prospects to verify a possible
  dark matter hint in cosmic antiprotons with antideuterons and antihelium}'',
  \href{http://dx.doi.org/10.1103/PhysRevD.97.103011}{{\em Phys. Rev.}
  {\bfseries D97} (2018) 103011},
\href{http://arxiv.org/abs/1711.08465}{{\ttfamily arXiv:1711.08465
  [astro-ph.HE]}}.

\bibitem{Bellini:2018epz}
F.~Bellini and A.~P. Kalweit, ``{Testing coalescence and statistical-thermal
  production scenarios for (anti-)(hyper-)nuclei and exotic QCD objects at LHC
  energies}'', \href{http://dx.doi.org/10.1103/PhysRevC.99.054905}{{\em Phys.
  Rev.} {\bfseries C99} (2019) 054905},
\href{http://arxiv.org/abs/1807.05894}{{\ttfamily arXiv:1807.05894 [hep-ph]}}.

\bibitem{Oliinychenko:2018ugs}
D.~Oliinychenko, L.-G. Pang, H.~Elfner, and V.~Koch, ``{Microscopic study of
  deuteron production in PbPb collisions at $\sqrt{s} = 2.76 TeV$ via
  hydrodynamics and a hadronic afterburner}'',
  \href{http://dx.doi.org/10.1103/PhysRevC.99.044907}{{\em Phys. Rev. C}
  {\bfseries 99} (2019) 044907},
  \href{http://arxiv.org/abs/1809.03071}{{\ttfamily arXiv:1809.03071
  [hep-ph]}}.

\bibitem{Andronic:2017pug}
A.~Andronic, P.~Braun-Munzinger, K.~Redlich, and J.~Stachel, ``{Decoding the
  phase structure of QCD via particle production at high energy}'',
  \href{http://dx.doi.org/10.1038/s41586-018-0491-6}{{\em Nature} {\bfseries
  561} (2018) 321--330}, \href{http://arxiv.org/abs/1710.09425}{{\ttfamily
  arXiv:1710.09425 [nucl-th]}}.

\bibitem{ALICE:2020ibs}
{\bfseries ALICE} Collaboration, S.~Acharya {\em et~al.}, ``{Search for a
  common baryon source in high-multiplicity pp collisions at the LHC}'',
  \href{http://dx.doi.org/10.1016/j.physletb.2020.135849}{{\em Phys. Lett. B}
  {\bfseries 811} (2020) 135849},
  \href{http://arxiv.org/abs/2004.08018}{{\ttfamily arXiv:2004.08018
  [nucl-ex]}}.

\bibitem{Bellini:2020cbj}
F.~Bellini, K.~Blum, A.~P. Kalweit, and M.~Puccio, ``{Examination of
  coalescence as the origin of nuclei in hadronic collisions}'',
  \href{http://dx.doi.org/10.1103/PhysRevC.103.014907}{{\em Phys. Rev. C}
  {\bfseries 103} (2021) 014907},
  \href{http://arxiv.org/abs/2007.01750}{{\ttfamily arXiv:2007.01750
  [nucl-th]}}.

\bibitem{Davis:2005mb}
D.~H. Davis, ``{50 years of hypernuclear physics. I. The early experiments}'',
  \href{http://dx.doi.org/10.1016/j.nuclphysa.2005.01.002}{{\em Nucl. Phys. A}
  {\bfseries 754} (2005) 3--13}.

\bibitem{Adam:2019phl}
{\bfseries STAR} Collaboration, J.~Adam {\em et~al.}, ``{Measurement of the
  mass difference and the binding energy of the hypertriton and
  antihypertriton}'', \href{http://dx.doi.org/10.1038/s41567-020-0799-7}{{\em
  Nature Phys.} {\bfseries 16} (2020) 409--412},
  \href{http://arxiv.org/abs/1904.10520}{{\ttfamily arXiv:1904.10520
  [hep-ex]}}.

\bibitem{Nemura:1999qp}
H.~Nemura, Y.~Suzuki, Y.~Fujiwara, and C.~Nakamoto, ``{Study of light Lambda
  and Lambda-Lambda hypernuclei with the stochastic variational method and
  effective Lambda N potentials}'',
  \href{http://dx.doi.org/10.1143/PTP.103.929}{{\em Prog. Theor. Phys.}
  {\bfseries 103} (2000) 929--958},
  \href{http://arxiv.org/abs/nucl-th/9912065}{{\ttfamily
  arXiv:nucl-th/9912065}}.

\bibitem{Hildenbrand:2019sgp}
F.~Hildenbrand and H.~W. Hammer, ``{Three-Body Hypernuclei in Pionless
  Effective Field Theory}'',
  \href{http://dx.doi.org/10.1103/PhysRevC.100.034002}{{\em Phys. Rev. C}
  {\bfseries 100} (2019) 034002},
  \href{http://arxiv.org/abs/1904.05818}{{\ttfamily arXiv:1904.05818
  [nucl-th]}}. [Erratum: Phys.Rev.C 102, 039901 (2020)].

\bibitem{Abelev:2014pja}
{\bfseries ALICE} Collaboration, B.~Abelev {\em et~al.}, ``{Freeze-out radii
  extracted from three-pion cumulants in pp, p\textendash{}Pb and
  Pb\textendash{}Pb collisions at the LHC}'',
  \href{http://dx.doi.org/10.1016/j.physletb.2014.10.034}{{\em Phys. Lett. B}
  {\bfseries 739} (2014) 139--151},
  \href{http://arxiv.org/abs/1404.1194}{{\ttfamily arXiv:1404.1194 [nucl-ex]}}.

\bibitem{Adam:2015pya}
{\bfseries ALICE} Collaboration, J.~Adam {\em et~al.}, ``{Two-pion femtoscopy
  in p-Pb collisions at $\sqrt{s_{\rm NN}}=5.02$ TeV}'',
  \href{http://dx.doi.org/10.1103/PhysRevC.91.034906}{{\em Phys. Rev. C}
  {\bfseries 91} (2015) 034906},
  \href{http://arxiv.org/abs/1502.00559}{{\ttfamily arXiv:1502.00559
  [nucl-ex]}}.

\bibitem{Aamodt:2008zz}
{\bfseries ALICE} Collaboration, K.~Aamodt {\em et~al.}, ``{The ALICE
  experiment at the CERN LHC}'',
  \href{http://dx.doi.org/10.1088/1748-0221/3/08/S08002}{{\em JINST} {\bfseries
  3} (2008) S08002}.

\bibitem{Abelev:2014ffa}
{\bfseries ALICE} Collaboration, B.~Abelev {\em et~al.}, ``{Performance of the
  ALICE Experiment at the CERN LHC}'',
  \href{http://dx.doi.org/10.1142/S0217751X14300440}{{\em Int. J. Mod. Phys.}
  {\bfseries A29} (2014) 1430044},
\href{http://arxiv.org/abs/1402.4476}{{\ttfamily arXiv:1402.4476 [nucl-ex]}}.

\bibitem{Abbas:2013taa}
{\bfseries ALICE} Collaboration, E.~Abbas {\em et~al.}, ``{Performance of the
  ALICE VZERO system}'',
  \href{http://dx.doi.org/10.1088/1748-0221/8/10/P10016}{{\em JINST} {\bfseries
  8} (2013) P10016},
\href{http://arxiv.org/abs/1306.3130}{{\ttfamily arXiv:1306.3130 [nucl-ex]}}.

\bibitem{Abelev:2014epa}
{\bfseries ALICE} Collaboration, B.~B. Abelev {\em et~al.}, ``{Measurement of
  visible cross sections in proton-lead collisions at $\sqrt{s_{\rm NN}}$ =
  5.02 TeV in van der Meer scans with the ALICE detector}'',
  \href{http://dx.doi.org/10.1088/1748-0221/9/11/P11003}{{\em JINST} {\bfseries
  9} (2014) P11003}, \href{http://arxiv.org/abs/1405.1849}{{\ttfamily
  arXiv:1405.1849 [nucl-ex]}}.

\bibitem{Aamodt:2010aa}
{\bfseries ALICE} Collaboration, K.~Aamodt {\em et~al.}, ``{Alignment of the
  ALICE Inner Tracking System with cosmic-ray tracks}'',
  \href{http://dx.doi.org/10.1088/1748-0221/5/03/P03003}{{\em JINST} {\bfseries
  5} (2010) P03003},
\href{http://arxiv.org/abs/1001.0502}{{\ttfamily arXiv:1001.0502
  [physics.ins-det]}}.

\bibitem{Alme:2010ke}
J.~Alme {\em et~al.}, ``{The ALICE TPC, a large 3-dimensional tracking device
  with fast readout for ultra-high multiplicity events}'',
  \href{http://dx.doi.org/10.1016/j.nima.2010.04.042}{{\em Nucl. Instrum.
  Meth.} {\bfseries A622} (2010) 316--367},
\href{http://arxiv.org/abs/1001.1950}{{\ttfamily arXiv:1001.1950
  [physics.ins-det]}}.

\bibitem{Adam:2015yta}
{\bfseries ALICE} Collaboration, J.~Adam {\em et~al.},
  ``{$^{3}_{\Lambda}\mathrm H$ and $^{3}_{\bar{\Lambda}} \overline{\mathrm H}$
  production in Pb-Pb collisions at $\sqrt{s_{\rm NN}} =$ 2.76 TeV}'',
  \href{http://dx.doi.org/10.1016/j.physletb.2016.01.040}{{\em Phys. Lett. B}
  {\bfseries 754} (2016) 360--372},
  \href{http://arxiv.org/abs/1506.08453}{{\ttfamily arXiv:1506.08453
  [nucl-ex]}}.

\bibitem{Acharya:2019qcp}
{\bfseries ALICE} Collaboration, S.~Acharya {\em et~al.},
  ``{$^3_\Lambda\mathrm{H}$ and $^3_{\bar{\Lambda}}\mathrm{\overline{H}}$
  lifetime measurement in Pb-Pb collisions at $\sqrt{s_{\mathrm{NN}}} = $ 5.02
  TeV via two-body decay}'',
  \href{http://dx.doi.org/10.1016/j.physletb.2019.134905}{{\em Phys. Lett. B}
  {\bfseries 797} (2019) 134905},
  \href{http://arxiv.org/abs/1907.06906}{{\ttfamily arXiv:1907.06906
  [nucl-ex]}}.

\bibitem{10.1145/2939672.2939785}
T.~Chen and C.~Guestrin,
  \href{http://dx.doi.org/10.1145/2939672.2939785}{``Xgboost: A scalable tree
  boosting system'',} in {\em Proceedings of the 22nd ACM SIGKDD International
  Conference on Knowledge Discovery and Data Mining}, KDD '16, p.~785–794.
\newblock Association for Computing Machinery, New York, NY, USA, 2016.
\newblock \url{https://doi.org/10.1145/2939672.2939785}.

\bibitem{ATLAS:2018mme}
{\bfseries ATLAS} Collaboration, M.~Aaboud {\em et~al.}, ``{Observation of
  Higgs boson production in association with a top quark pair at the LHC with
  the ATLAS detector}'',
  \href{http://dx.doi.org/10.1016/j.physletb.2018.07.035}{{\em Phys. Lett. B}
  {\bfseries 784} (2018) 173--191},
  \href{http://arxiv.org/abs/1806.00425}{{\ttfamily arXiv:1806.00425
  [hep-ex]}}.

\bibitem{barioglio_luca_2021_5734093}
L.~Barioglio, F.~Catalano, M.~Concas, P.~Fecchio, F.~Grosa, F.~Mazzaschi, and
  M.~Puccio, ``hipe4ml/hipe4ml'', Nov., 2021.
\newblock \url{https://doi.org/10.5281/zenodo.5734093}.

\bibitem{Wang:1991hta}
X.-N. Wang and M.~Gyulassy, ``{HIJING: A Monte Carlo model for multiple jet
  production in pp, pA and AA collisions}'',
  \href{http://dx.doi.org/10.1103/PhysRevD.44.3501}{{\em Phys. Rev. D}
  {\bfseries 44} (1991) 3501--3516}.

\bibitem{Agostinelli:2002hh}
{\bfseries GEANT4} Collaboration, S.~Agostinelli {\em et~al.}, ``{GEANT4--a
  simulation toolkit}'',
  \href{http://dx.doi.org/10.1016/S0168-9002(03)01368-8}{{\em Nucl. Instrum.
  Meth. A} {\bfseries 506} (2003) 250--303}.

\bibitem{Cranmer:2000du}
K.~S. Cranmer, ``{Kernel estimation in high-energy physics}'',
  \href{http://dx.doi.org/10.1016/S0010-4655(00)00243-5}{{\em Comput. Phys.
  Commun.} {\bfseries 136} (2001) 198--207},
  \href{http://arxiv.org/abs/hep-ex/0011057}{{\ttfamily arXiv:hep-ex/0011057}}.

\bibitem{Verkerke:2003ir}
W.~Verkerke and D.~P. Kirkby, ``{The RooFit toolkit for data modeling}'', {\em
  eConf} {\bfseries C0303241} (2003) MOLT007,
  \href{http://arxiv.org/abs/physics/0306116}{{\ttfamily
  arXiv:physics/0306116}}.

\bibitem{Cowan:2010js}
G.~Cowan, K.~Cranmer, E.~Gross, and O.~Vitells, ``{Asymptotic formulae for
  likelihood-based tests of new physics}'',
  \href{http://dx.doi.org/10.1140/epjc/s10052-011-1554-0}{{\em Eur. Phys. J. C}
  {\bfseries 71} (2011) 1554}, \href{http://arxiv.org/abs/1007.1727}{{\ttfamily
  arXiv:1007.1727 [physics.data-an]}}. [Erratum: Eur.Phys.J.C 73, 2501 (2013)].

\bibitem{Kamada:1997rv}
H.~Kamada, J.~Golak, K.~Miyagawa, H.~Witala, and W.~Gloeckle, ``{Pi mesonic
  decay of the hypertriton}'',
  \href{http://dx.doi.org/10.1103/PhysRevC.57.1595}{{\em Phys. Rev. C}
  {\bfseries 57} (1998) 1595--1603},
  \href{http://arxiv.org/abs/nucl-th/9709035}{{\ttfamily
  arXiv:nucl-th/9709035}}.

\bibitem{Schnedermann:1993ws}
E.~Schnedermann, J.~Sollfrank, and U.~W. Heinz, ``{Thermal phenomenology of
  hadrons from 200-A/GeV S+S collisions}'',
  \href{http://dx.doi.org/10.1103/PhysRevC.48.2462}{{\em Phys. Rev. C}
  {\bfseries 48} (1993) 2462--2475},
  \href{http://arxiv.org/abs/nucl-th/9307020}{{\ttfamily
  arXiv:nucl-th/9307020}}.

\bibitem{Abelev:2013haa}
{\bfseries ALICE} Collaboration, B.~Abelev {\em et~al.}, ``{Multiplicity
  Dependence of Pion, Kaon, Proton and Lambda Production in p-Pb Collisions at
  $\sqrt{s_{NN}}$ = 5.02 TeV}'',
  \href{http://dx.doi.org/10.1016/j.physletb.2013.11.020}{{\em Phys. Lett. B}
  {\bfseries 728} (2014) 25--38},
  \href{http://arxiv.org/abs/1307.6796}{{\ttfamily arXiv:1307.6796 [nucl-ex]}}.

\bibitem{Evlanov:1998py}
M.~V. Evlanov, A.~M. Sokolov, V.~K. Tartakovsky, S.~A. Khorozov, and
  Y.~Lukstinsh, ``{Interaction of hypertritons with nuclei at high-energies}'',
  \href{http://dx.doi.org/10.1016/S0375-9474(98)00116-X}{{\em Nucl. Phys. A}
  {\bfseries 632} (1998) 624--632}.

\bibitem{Vovchenko:2019pjl}
V.~Vovchenko and H.~Stoecker, ``{Thermal-FIST: A package for heavy-ion
  collisions and hadronic equation of state}'',
  \href{http://dx.doi.org/10.1016/j.cpc.2019.06.024}{{\em Comput. Phys.
  Commun.} {\bfseries 244} (2019) 295--310},
  \href{http://arxiv.org/abs/1901.05249}{{\ttfamily arXiv:1901.05249
  [nucl-th]}}.

\bibitem{Zhang:2009ba}
S.~Zhang, J.~H. Chen, H.~Crawford, D.~Keane, Y.~G. Ma, and Z.~B. Xu,
  ``{Searching for onset of deconfinement via hypernuclei and
  baryon-strangeness correlations}'',
  \href{http://dx.doi.org/10.1016/j.physletb.2010.01.034}{{\em Phys. Lett. B}
  {\bfseries 684} (2010) 224--227},
  \href{http://arxiv.org/abs/0908.3357}{{\ttfamily arXiv:0908.3357 [nucl-ex]}}.

\bibitem{Gyulassy:1994ew}
M.~Gyulassy and X.-N. Wang, ``{HIJING 1.0: A Monte Carlo program for parton and
  particle production in high-energy hadronic and nuclear collisions}'',
  \href{http://dx.doi.org/10.1016/0010-4655(94)90057-4}{{\em Comput. Phys.
  Commun.} {\bfseries 83} (1994) 307},
  \href{http://arxiv.org/abs/nucl-th/9502021}{{\ttfamily
  arXiv:nucl-th/9502021}}.

\bibitem{Pierog:2013ria}
T.~Pierog, I.~Karpenko, J.~M. Katzy, E.~Yatsenko, and K.~Werner, ``{EPOS LHC:
  Test of collective hadronization with data measured at the CERN Large Hadron
  Collider}'', \href{http://dx.doi.org/10.1103/PhysRevC.92.034906}{{\em Phys.
  Rev. C} {\bfseries 92} (2015) 034906},
  \href{http://arxiv.org/abs/1306.0121}{{\ttfamily arXiv:1306.0121 [hep-ph]}}.

\bibitem{Roesler:2000he}
S.~Roesler, R.~Engel, and J.~Ranft,
  \href{http://dx.doi.org/10.1007/978-3-642-18211-2_166}{``{The Monte Carlo
  event generator DPMJET-III}'',} in {\em {International Conference on Advanced
  Monte Carlo for Radiation Physics, Particle Transport Simulation and
  Applications (MC 2000)}}, pp.~1033--1038.
\newblock Springer, Berlin, Dec, 2000.
\newblock \href{http://arxiv.org/abs/hep-ph/0012252}{{\ttfamily
  arXiv:hep-ph/0012252}}.

\bibitem{ALICE:2012xs}
{\bfseries ALICE} Collaboration, B.~Abelev {\em et~al.}, ``{Pseudorapidity
  density of charged particles in $p$ + Pb collisions at $\sqrt{s_{NN}}=5.02$
  TeV}'', \href{http://dx.doi.org/10.1103/PhysRevLett.110.032301}{{\em Phys.
  Rev. Lett.} {\bfseries 110} (2013) 032301},
  \href{http://arxiv.org/abs/1210.3615}{{\ttfamily arXiv:1210.3615 [nucl-ex]}}.

\bibitem{Adamczyk:2017buv}
{\bfseries STAR} Collaboration, L.~Adamczyk {\em et~al.}, ``{Measurement of the
  $^3_{\Lambda}$H lifetime in Au+Au collisions at the BNL Relativistic Heavy
  Ion Collider}'', \href{http://dx.doi.org/10.1103/PhysRevC.97.054909}{{\em
  Phys. Rev. C} {\bfseries 97} (2018) 054909},
  \href{http://arxiv.org/abs/1710.00436}{{\ttfamily arXiv:1710.00436
  [nucl-ex]}}.

\bibitem{Cleymans:2020fsc}
J.~Cleymans, P.~M. Lo, K.~Redlich, and N.~Sharma, ``{Multiplicity dependence of
  (multi)strange baryons in the canonical ensemble with phase shift
  corrections}'', \href{http://dx.doi.org/10.1103/PhysRevC.103.014904}{{\em
  Phys. Rev. C} {\bfseries 103} (2021) 014904},
  \href{http://arxiv.org/abs/2009.04844}{{\ttfamily arXiv:2009.04844
  [hep-ph]}}.

\bibitem{ALICE-PUBLIC-2020-005}
{\bfseries ALICE} Collaboration, ``{Future high-energy pp programme with
  ALICE}'', 2020.
\newblock \url{http://cds.cern.ch/record/2724925}.

\end{thebibliography}\endgroup
